\documentclass[prb,twocolumn,amsmath,showpacs]{revtex4}
\input {epsf.sty}
\usepackage{epsfig}
\usepackage{amsmath,amsthm,amssymb,latexsym,amsfonts}

\begin{document}
\title{Impurity Effects of Aerogel in Superfluid $^{3}$He}

\author{W.P. Halperin, H. Choi, J.P. Davis, and J. Pollanen}

\affiliation{Department of Physics and Astronomy, \\Northwestern  University, Evanston, IL 60208,
USA}
\date{Version \today}


\begin{abstract}The discovery of superfluid $^{3}$He in high porosity silica aerogels, and
subsequent experimental and theoretical work, have led to a better general understanding of
quasiparticle scattering from impurities  in unconventional pairing systems.  It is immensely
helpful for understanding impurity effects in the case of superfluid
$^{3}$He that the structure of its order parameter is
well-established. An overview of impurity effects is
presented with emphasis on those experiments which have a quantitative interpretation
in terms of theoretical models for homogeneous and inhomogeneous scattering. The latter can
account successfully for most experimental results. 
\end{abstract}

\maketitle

\vspace{11pt}
\section{Introduction}
Since its discovery\cite{Osh97} in 1971, superfluid $^3$He has been  influential for investigations
of quantum condensed systems including exotic and unconventional superconductivity,\cite{Ann04}
quantum gases,\cite{Ann04,Leg06} neutron stars,\cite{Pin85} and  systems with complex spontaneous
symmetry breaking, as for example in quark matter.\cite{Raj01}  The
impact of superfluid
$^3$He research in condensed matter physics has been most closely related to unconventional
superconducting pairing condensations, high temperature superconductivity and certain heavy
fermion superconductors.\cite{Ann04}  In these cases the effects of quasiparticle scattering, either
from chemical impurities or material defects, can be significant\cite{Bal06} and yet they can be
difficult to interpret.   For these exotic superconductors, one of the difficulties  is that the
 order parameter of the pure system may not be established. It is
helpful for developing basic ideas to compare with
$^3$He where the order parameter structure is very well-known and where there are high resolution
experimental techniques to probe both spin and orbital dependences of the
condensed state.  Consequently,
$^3$He is an excellent forum for investigating impurity effects; however, it was not known until
 recently how to systematically introduce such impurities into $^3$He, the
purest of any known material. 

About a dozen years ago groups at Cornell University  and Northwestern
University investigated samples of 98\% porosity silica aerogel samples provided by Moses Chan from
Pennsylvania State University, to see what might be the behavior at low temperatures of
$^3$He imbibed  into these highly porous structures.  Both groups reported
evidence for sharp onsets of superfluidity,\cite{Por95,Spr95,Spr96} albeit at
transition temperatures significantly reduced compared with the pure superfluid, as shown in the
phase diagram of the transition temperature, $T_{ca}$, versus pressure in Fig.~1.  A flood of
questions then arose: What was the new pressure-temperature phase diagram?  Is it reproducible from
one aerogel sample to the next? To what extent is the order parameter suppressed and does it track
the reduction in the transition temperature?  What are the equilibrium phases? Are there new physical
phenomena different from the pure system?  Concerning the last question,  there is currently
substantial interest in a metastable equal-spin-pairing state, an
$A$-like phase, that appears on cooling through the normal-to-superfluid transition. It was 
suggested by Vicente {\it et al.}\cite{Vic05} that the origin of this phase might be associated
with anisotropic scattering. Subsequently, some theoretical work\cite{Aoy06} and  a number of
experiments\cite{Dav06,Dav08,Kun07,Dmi08,Elb08,Sat08} have investigated
$^3$He in aerogels having global anisotropy, suggesting new order parameter structures and perhaps
new kinds of superfluid phases.  These questions and others have been the subject of intense
activity.  

%%%%%%%%%%%%%%%%%%%%%%%%%%%%%%%%%%%%%%%%%%%%%%

\begin{figure}[b]
%%%%%%%%%%%%%%%%%   F I G U R E  1   %%%%%%%%%%%%%%%%%%
\centerline{\includegraphics[width=2.8in]{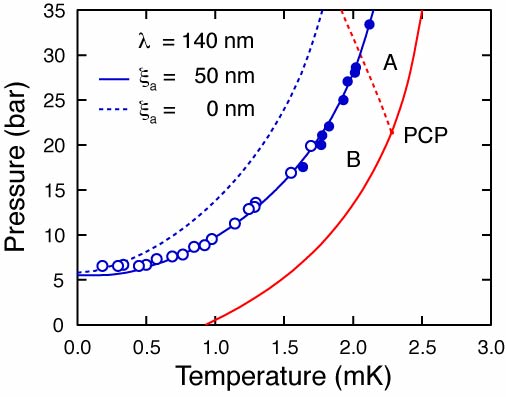}}
%%%%%%%%%%%%%%%%%%%%%%%%%%%%%%%%%%%%%%%%%%%%%  
\begin{minipage}{0.93\hsize}
\caption {Phase diagram for superfluid $^3$He in  two different samples of 98\% aerogel. The 
known superfluid phases of pure $^3$He are labeled A and B meeting with the normal phase at a
polycritical point, PCP.  The open circles are measurements from the Cornell group\cite{Mat97} and
the closed circles from the Northwestern group\cite{Ger01,Ger02} both for nominally 98\% porous
aerogels made by Norbert Mulders at the University of Delaware. The blue curves are from theoretical
models discussed in the text.}
\end{minipage}
\end{figure}  

In this article we concentrate mainly on a number of experimental results  which have been
interpreted quantitatively in terms of theoretical models for impurity scattering.  As such our
review is not a comprehensive treatment of superfluid $^3$He in aerogel, although it is
complementary to an introductory review by Halperin and Sauls.\cite{Hal04} We outline the principal
features of superfluid
$^3$He in aerogel; we introduce the GL-theory for a
$p$-wave superfluid and its adaption to theoretical models for impurity scattering; and we discuss
the experimental results in this context.  Lastly, we discuss the metastable
anisotropic phase, and some of the recent work on anisotropic aerogels that attempt to uncover the
origin of this fascinating phenomenon of metastability between different $p$-wave superfluid states.

\begin{figure}[t]
%%%%%%%%%%%%%%%%%   F I G U R E  2   %%%%%%%%%%%%%%%%%%
\centerline{\includegraphics[width=2.8in]{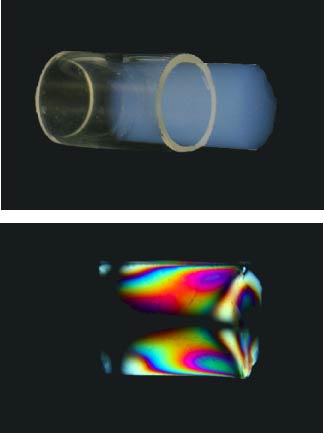}}
%%%%%%%%%%%%%%%%%%%%%%%%%%%%%%%%%%%%%%%%%%%%%
\begin{minipage}{0.93\hsize}
\caption {High porosity silica aerogel made at Northwestern. The upper image shows a sample after
supercritical drying partially removed from the glass tube in which it was formed.  The lower image
is from exposure to white light passing through crossed polarizers, an example of an
inhomogeneous aerogel with both density variations and anisotropy that can result from
abnormally high concentrations of catalyst used during synthesis.  Both samples have 98\%
porosity. In contrast, a homogeneous aerogel would not transmit any light and the image would be
black.}
\end{minipage}
\end{figure}
\noindent

\begin{figure}[t]
%%%%%%%%%%%%%%%%%   F I G U R E  3   %%%%%%%%%%%%%%%%%%
\centerline{\includegraphics[width=2.8in]{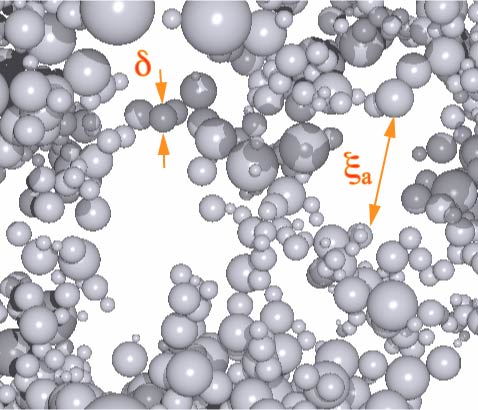}}
%%%%%%%%%%%%%%%%%%%%%%%%%%%%%%%%%%%%%%%%%%%%%
\begin{minipage}{0.93\hsize}
\caption {Perspective view of a DLCA calculation of a 98\% aerogel.  The principal length scales
can be seen including the SiO$_2$ particles $\approx 3$ nm in diameter and their
correlation length $\xi_{a}\approx 40$ nm.  The transport mean free path is much longer
($\lambda >> \xi_{a}$) given the open nature of the structure, $\lambda\approx 180$ nm. This
simulation was made by Tom Lippman at Northwestern University. Simulations have been
performed by Porto and Parpia\cite{Por99} and Haard {\it et al.}\cite{Haa00,Haa01}}
\end{minipage}
\end{figure}

\noindent
\vspace{11pt}
\section{High Porosity Silica Aerogel}

Highly porous silica aerogel is itself a subject
of widespread interest to diverse applications including the collection of
cosmic interstellar particles, thermal insulation, catalysis, particle detectors, radiation
absorbers, liquid crystal phase transitions, and applications to quantum fluids and solids, where
the present topic is but one example.  See Pollanen {\it et al.}\cite{Pol08} and references therein
concerning these applications. In Fig.~2 we show a typical sample of 98\% porous silica aerogel
grown at Northwestern University partly extracted from the glass tube in which it was produced. 
Under some synthesis conditions, including high concentrations of base catalyst,  the aerogels
will exhibit anisotropy that can be easily visualized with white light and crossed polarizers as
shown in the colorful picture in the second panel of Fig.~2.  However, the synthesis can be
controlled such that the anisotropy is either globally axial, or completely avoided
altogether.\cite{Pol08} In the latter case, an image  with
polarized light of a homogeneous and isotropic aerogel sample is black for any angle of presentation
with respect to the polarizers. In the case of homogeneous axial anisotropy the image is black only
for the anisotropy axis (optical axis) parallel to either of the polarizers. It is important for
applications to $^3$He that the structure of the aerogel  be well-characterized using optical
birefringence methods in combination with small angle x-ray measurements (SAXS) as discussed by
Pollanen {\it et al.}\cite{Pol08} We have grown isotropic silica aerogels by the one-step method in
the range of 95 to 99\% porosity and 98\% porosity using the two-step method; in both cases we use a
technique of rapid supercritical extraction\cite{Poc96} during the drying stage.

\begin{figure}[t]
%%%%%%%%%%%%%%%%%   F I G U R E  4   %%%%%%%%%%%%%%%%%%
\centerline{\includegraphics[width=2.8in]{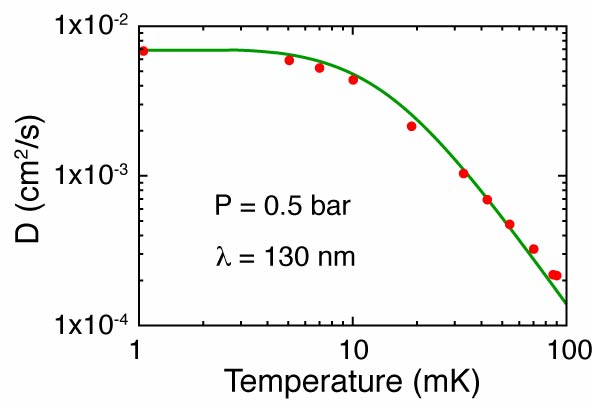}}
%%%%%%%%%%%%%%%%%%%%%%%%%%%%%%%%%%%%%%%%%%%%%
\begin{minipage}{0.93\hsize}
\caption {Diffusion coefficient for normal $^3$He determined from NMR by the Grenoble
group.\cite{Bun05,Sau05}  The leveling off of diffusion with decreasing temperature indicates the
transition from inelastic to elastic scattering from which $\lambda = 130$ nm was obtained from
consideration of elastic scattering in the normal Fermi liquid.}
\end{minipage}
\end{figure}
\noindent

Additionally, numerical simulations can be very helpful for understanding aerogel.  They
are  generally based on diffusion limited cluster aggregation (DLCA) algorithms.  These
methods\cite{Haa00,Haa01,Por99} generate fractal like structures such as shown in Fig.~3, that permit
calculation of the x-ray scattering crossection and the transport (ballistic) mean free path,
$\lambda$. This may be the best way to correlate measured transport coefficients in quantum fluids
with microscopic characterization using SAXS.  The simulation in
Fig.~3 is a perspective view of a calculation performed by Tom Lippman at Northwestern University
using DLCA.  The  principal length scales of the structure are indicated: the  SiO$_2$ particle size
of
$\delta \approx 3$ nm, and the particle-particle correlation length $\xi_{a}\approx 40$ nm. 

Ideally
$\lambda$ should be determined directly from transport experiments in the normal fluid:
acoustic attenuation,\cite{Nom00} thermal conductivity,\cite{Fis01} or diffusion
measurements.\cite{Bun05,Sau05}  The analysis of diffusion measurements performed by the Grenoble
group\cite{Sau05} are shown in Fig.~4 for a 98\% aerogel giving, $\lambda = 130$ nm. It is
noteworthy that the transport mean free path $\lambda$ determined in the normal state from diffusion,
matches well with the corresponding parameter obtained from analysis of superfluid $^3$He properties
using the inhomogeneous scattering theory.

\begin{figure}[t]
%%%%%%%%%%%%%%%%%   F I G U R E  5   %%%%%%%%%%%%%%%%%%
\centerline{\includegraphics[width=2.8in]{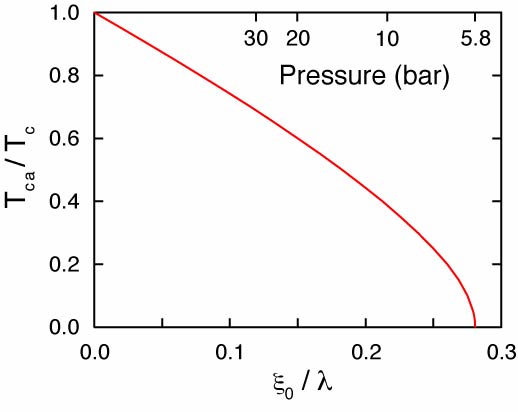}}
%%%%%%%%%%%%%%%%%%%%%%%%%%%%%%%%%%%%%%%%%%%%%
\begin{minipage}{0.93\hsize}
\caption {Reduction of the transition temperature by impurity scattering as a function of 
the pair breaking parameter, $x =
\xi_{0}/\lambda$, according to Abrikosov
and Gorkov.\cite{Abr61}  It can be easily seen that this behavior, to first order, accounts for the
phase diagram in Fig.~1 since a clockwise rotation of the above figure by $90^{\circ}$ maps onto
the phase diagram,  as suggested by the upper-scale for the pressure in this figure with $\lambda =
140$ nm.  The depairing parameter depends on the pressure through the pressure dependence of the
coherence length, $\xi_{0}$.  Critical depairing is reached near $P = 6$ bar where $x = 0.28$.}
\end{minipage}
\end{figure}
\noindent

\vspace{11pt}
\section{Impurity Scattering from Silca Aerogel}

The effect of magnetic impurity scattering on the superconducting transition of an $s$-wave
superconductor was calculated by Abrikosov and Gorkov (AG)\cite{Abr61} and extended by
Tsuneto\cite{Tsu62} and Larkin\cite{Lar65} to non-$s$-wave pairing, where any form of impurity
scattering has the effect of depairing.  To adapt this theory to $p$-wave superfluids\cite{Thu98}
we define a pair breaking parameter,
$x$, to be the ratio of the coherence length at zero temperature of pure
$^3$He to the transport mean free path, $x =
\xi_{0}/\lambda$ where $\xi_{0} = \hbar v_{F}/(2\pi k_{B}T_{c})$.  According to the
AG-theory,\cite{Abr61} the dependence of
$T_{ca}/T_{c}$ on $x$, is given in Fig.~5, where
$T_{c}$ is the transition temperature in the absence of elastic scattering.  In the clean
limit there is a linear relationship.  For large scattering rates there is a critical value of $x$
beyond which the superconducting state is not stable.  These  features are  qualitatively manifest in
the observed phase diagram for superfluid $^3$He in aerogel shown in Fig.~1. Here $\lambda$ is fixed
by the silica structure while $\xi_{0}$ varies between $16$ nm, at high pressure, to $77$
nm at zero  pressure. Matsumoto {\it et al.}\cite{Mat97} discovered a critical pressure for
superfluidity at 
$P \approx 6$ bar in a $98\%$ aerogel which
corresponds to the critical value for the depairing parameter in Fig.~5, predicted by Abrikosov
and Gorkov.\cite{Abr61}

\begin{figure}[b]
%%%%%%%%%%%%%%%%%   F I G U R E  6   %%%%%%%%%%%%%%%%%%
\centerline{\includegraphics[width=2.8in]{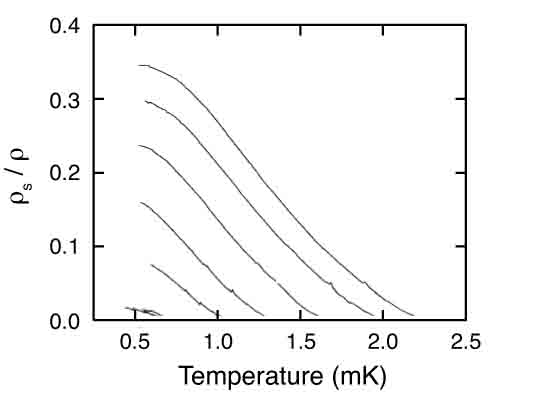}}
%%%%%%%%%%%%%%%%%%%%%%%%%%%%%%%%%%%%%%%%%%%%%
\begin{minipage}{0.93\hsize}
\caption {Superfluid density and transition temperature for superfluidity from torsional
oscillator measurements from the Cornell University group of Porto and Parpia.\cite{Por95}  The
pressures shown for these curves are, from lowest to highest: 3.4 5.0, 7.0, 10, 15, and 25 bar. This
sample showed less suppression of
$T_{ca}$ than that reported later by Matsumoto {\it et al.}\cite{Mat97} (Fig.~1.)}
\end{minipage}
\end{figure}
\noindent

\section{Indicators of T$_c$}

The first indication of a transition temperature for superfluidity of $^3$He in a 98\% aerogel was
found by Porto and Parpia\cite{Por95} from the period shift measurements of a torsional oscillator
that they interpreted as an unlocking of the superfluid component. They obtained
the superfluid density as shown in Fig.~6.  Shortly thereafter, Sprague {\it et
al.}\cite{Spr95,Spr96} found frequency shifts of the NMR spectrum, Fig.~7, which displayed a
sudden onset at a given temperature and were similarily interpreted. Transition temperatures
obtained later by these two groups are intercompared in the phase diagram in Fig.~1 indicating the
same degree of suppression.  It was clear from these first
experiments that the transition temperature is systematically reduced compared to that of the pure
superfluid, and this effect is more pronounced at lower pressures.  This is expected from
AG-theory.  But it was also clear that the low temperature limiting value for the superfluid
density, particularly at low pressure, was not 100\%. Similarily the magnitude of the NMR frequency
shifts was substantially reduced from values expected for pure
$^3$He.  In both cases it appears that the order parameter is  suppressed, but more than
expected from scaling of the transition temperature alone.  From the AG-theory, adapted to
$p$-wave superfluidity\cite{Thu98} with unitary
scattering, the suppression of these two basic superfluid characteristics should be the same.  In
Fig.~8 a comparison is made to demonstrate this point.  It can be seen that the transition
temperature for the superfluid in aerogel, relative to the pure superfluid, is reduced, but
systematically less so as compared to the amplitude of the order parameter,
$\Delta(T)$, in the Ginzburg-Landau limit. The square of the order
parameter amplitude near  $T_{c}$ is 

\begin{figure}[t]
%%%%%%%%%%%%%%%%%   F I G U R E  7   %%%%%%%%%%%%%%%%%%
\centerline{\includegraphics[width=2.8in]{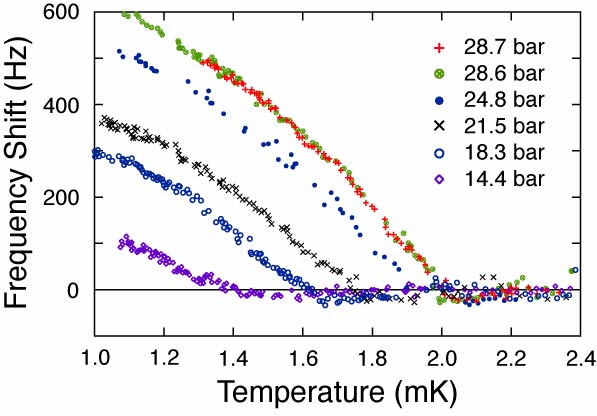}}
%%%%%%%%%%%%%%%%%%%%%%%%%%%%%%%%%%%%%%%%%%%%%
\begin{minipage}{0.93\hsize}
\caption {NMR frequency shifts measured by the Northwestern group\cite{Spr95,Spr96,Haa01} for
superfluid
$^3$He in a 98\% aerogel.  There is a clear indication of the
transition temperature. The magnitude of the frequency shifts at each pressure, corrected for
surface solid $^3$He, is reduced by a simple scale factor relative to the pure $A$-phase as
reported by Sprague {\it et al.}\cite{Spr95} and shown in Fig.~8. These authors identified an
equal-spin-pairing state, like the $A$-phase.}
\end{minipage}
\end{figure}
\noindent

\begin{figure}[b]
%%%%%%%%%%%%%%%%%   F I G U R E  8   %%%%%%%%%%%%%%%%%%
\centerline{\includegraphics[width=2.8in]{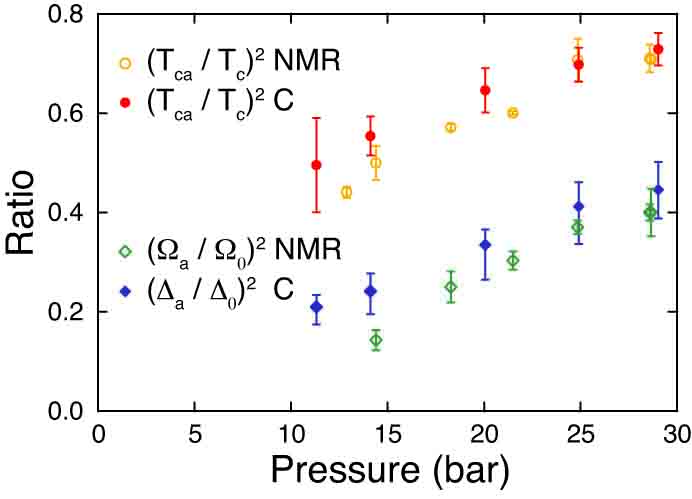}}
%%%%%%%%%%%%%%%%%%%%%%%%%%%%%%%%%%%%%%%%%%%%%
\begin{minipage}{0.93\hsize}
\caption {Suppression of the transition temperature and the amplitude of the order parameter as a
function of pressure.  The amplitude of the order parameter is determined from the NMR frequency
shifts and independently from the specific heat jump measurements of Choi {\it et al.}\cite{Cho04}
According to the homogeneous isotropic scattering model, with unitary
scattering,\cite{Thu98} the suppression factor for the transition temperature and the order parameter
amplitude should be the same, in contrast to the experiment.  Note that the 98\% aerogel
samples for specific heat and NMR experiments are slightly different from one another.}
\end{minipage}
\end{figure}
\noindent

\begin{equation}
\Delta^{2}(T) = {\left(\pi k_{B} T_{c}\right)^{2} {2 \over 3}}{\Delta
C \over C}{\left({T_{c} \over T} -1\right)}, 
\end{equation}
which can be directly determined from the specific heat  jump, $\Delta C/C$, such as measured by
Choi {\it et al.}\cite{Cho04} Additionally, measurements of the NMR frequency shift,
$\Delta\omega$, where
$\Delta^2
\propto \Delta\omega$, can be taken from Sprague {\it et
al.}\cite{Spr95,Spr96} to infer suppression of the order parameter. The two experiments, shown in
Fig. 8, are in excellent agreement, as will be discussed in subsequent sections.   However the
different suppression factors for the transition temperature and the order parameter indicate that
a strict application of the AG-theory, adapted to
$p$-wave pairing, the so-called homogeneous isotropic scattering model (HISM), does not suffice.

\section{GL-Theory and Experiment}

According to Ginzburg-Landau theory, just below the temperature for a second order
thermodynamic transition, the free energy can be represented phenomenologically as an expansion in 
terms of the order parameter.  In the case of superfluid $^3$He, the order 
parameter, $A$, is a complex,
second-rank tensor and the free energy can be expressed\cite{Leg75,Thu87,Vol90}  in
terms of its invariants.  Neglecting the dipole energy, this has the form,
\begin{eqnarray} F &=& -\alpha \mathrm{Tr}(AA^{\dagger})+ g_{z}H_{\mu}(AA^{\dagger})_{\mu\nu}H_{\nu}
+ \beta_{1}|\mathrm{Tr}(AA^{T})|^{2} \nonumber \\ && +\beta_{2}[\mathrm{Tr}(AA^{\dagger})]^{2} 
+\beta_{3}\mathrm{Tr}(AA^{T}(AA^{T})^{*}) \nonumber \\ && +
\beta_{4}\mathrm{Tr}((AA^{\dagger})^{2}) +\beta_{5}\mathrm{Tr}(AA^{\dagger}(AA^{\dagger})^{*})
\label{GL_free_energy}
\end{eqnarray} that we
have taken from Thuneberg\cite{Thu87} and the discussion by Choi {\it et al.}\cite{Cho07} The magnetic
field components are
$H_{\nu}$, and $A^{\dagger}$ and 
$A^{T}$ are the Hermitian conjugate and transpose of $A$. There are
 five fourth-order terms for which the coefficients, $\beta_{i}$,  determine the stable superfluid
states.  In the weak-coupling limit,
\begin{eqnarray} &\alpha={N(0)\over 3} \left({T\over T_{c}}-1\right),\\ &{\beta_{i}\over\beta_{0}} =
(-1, 2, 2, 2, -2),  i =1, ..., 5,\\ &\beta_{0}={7\zeta (3)\over 120 \pi^{2}} {N(0) \over
(k_{B}T_{c})^{2}},\\ &g_{z}={7\zeta(3)\over48\pi^{2}}N(0) \left( {\gamma_{0}\hbar \over
(1+F_{0}^{a})k_{B}T_{c}} \right)^{2},
\end{eqnarray} where the normal density of states at the Fermi energy is  $N(0)$, the gyromagnetic
ratio for $^3$He is $\gamma_{0}$, $k_{B}$ is the  Boltzmann constant, $F_{0}^{a}$ is a Fermi liquid
parameter determined from  magnetization measurements\cite{Vol90}  and
$\zeta(x)$ is the Riemann zeta function. For the weak-coupling superfluid the isotropic state,
called the $B$-phase, is always the most stable of all the $p$-wave states.  The existence of the
axial state at high pressure, known as the $A$-phase, is a result of strong-coupling effects
which are, to leading order, proportional to
$T_{c}/T_{F}$.\cite{Rai76}

The coefficient for the field coupling term of the order parameter, 
$g_{z}$, is determined by measuring the slope of the magnetization of $^{3}$He-$B$ in the limit 
approaching $T_{c}$.  The measurements to date for pure $^{3}$He are consistent with $g_{z}$ equal to
its weak-coupling value even at high pressure.\cite{Haa01,Cho07} This coefficient has the form,
\begin{equation}
g_{z}=g_{z}^{\mathrm wc}{{dm \over dt}\over{{dm \over dt}^{\mathrm wc}}}{{\beta_{B}} \over
{\beta_{B}^{\mathrm wc}}},
\end{equation}
where $m=M_{B}/M_{N}$, $M_{N}$ is the normal state magnetization, and $\beta_{B}$ is defined below. 
The superscript wc, which we use here and in the following,  indicates the weak-coupling limit.

It was shown by Greaves\cite{Gre76} that the rather small NMR $g$-shift\cite{Haa01,Cho07,Kyc94} of
the transverse NMR frequency in $^{3}$He-$B$ in the dipole-locked limit can be used to determine
$\beta_{345}$ in combination with $\hat g_{z} = g_{z}/g_{z}^{\mathrm wc}$.  

\begin{equation}
\frac{\beta_{345}}{\hat g_{z}}=\beta_{345}^{\mathrm wc} 
\frac{1}{(1+F_{0}^{a})^{2}} \left(
\frac{C_{N}}{\Delta C_{B}}\right) \frac{\nu_{B}^{2}}{1-t} 
\left( \frac{\hbar}{2 \pi k_{B} T_{c}} \right)^{2} \frac{1}{g}.
\label{Eq2}
\end{equation}
where $\nu_{B}$ is the longitudinal resonance frequency in the
$B$-phase measured by Rand {\it et al.},\cite{Ran94,Ran96} and
$t=T/T_{c}$. We use the standard notation, 
$\beta_{ij}=\beta_{i}+\beta_{j}$. The $g$-shift measurements are summarized by
Haard\cite{Haa01} and by Choi {\it et al.}\cite{Cho07}  

The specific heat,\cite{Gre86} $C_{N}$ and its 
jump $\Delta C_{A}$ and $\Delta C_{B}$ at
$T_{c}$ for the $A$ and $B$-phases, are related to
$\beta_{A}$ and $\beta_{B}$ through,
\begin{eqnarray} &{\Delta C_{A}}= 
\frac{\alpha'^{2}}{2\beta_{A}}, \beta_{A} \equiv \beta_{245} \\ &{\Delta C_{B}}=
\frac{\alpha'^{2}}{2\beta_{B}}, \beta_{B} 
\equiv \beta_{12} +
\frac{1}{3}\beta_{345},
\label{Eq4}
\end{eqnarray} where $\alpha\,' \equiv d\alpha/dT$.
The magnetic suppression\cite{Tan91} of $T_{AB}$, the transition from $A$ to $B$-phases is,
$g(\beta)$, 
\begin{equation} g(\beta) =
-{{\sqrt{1+(\beta_{B}/\beta_{A}-1)(1+{{2}\over{1-\beta_{12}/\beta_{B}}})}+1}\over{6(\beta_{B}/\beta_{A}-1})}.
\end{equation} Here $g(\beta)$ is defined by,

\begin{equation} 1- \frac{T_{AB}}{T_{c}} \equiv g(\beta)\left(B \over B_{0}\right)^2 +  O\left(B
\over B_{0}\right)^4,
\end{equation} where $B_{0}^2 = N(0)/6g_{z}$. Then finally $\beta_{5}$ can be determined by
measuring the asymmetry ratio\cite{Isr84} of the $A_{1}$-$A_{2}$ splitting in a high magnetic field,
$r$,

\begin{equation} r \equiv  \frac{T_{A1}-T_{c}}{T_{c}-T_{A2}} = -\frac{\beta_{5}}{\beta_{245}}.
\end{equation}

These strong-coupling effects
on the
$\beta$-parameters are determined by experiment. Recently a summary of these parameters for pure
$^3$He was given by Choi {\it et al.}\cite{Cho07} as well as their analysis for 98\% porosity
aerogel.  In fact there are only four independent experiments that relate to the $\beta$-parameters
in pure $^3$He.   Choi {\it et al.}\cite{Cho07} have gone one step further taking advantage of
theoretical calculations of Sauls and Serene\cite{Sau81} to propose one additional constraint based
on the calculated pressure dependence of these parameters. They obtain all five
$\beta$-parameters with reasonable confidence and correspondingly extend these to 98\% aerogel. Their
results are reproduced in Table 1 for a particular set of scattering parameters.  An important aspect
of superfluid $^3$He in aerogel is that suppression of the transition temperature by elastic
scattering reduces strong-coupling effects as compared to pure $^{3}$He.

\section{Homogeneous and Inhomogeneous Scattering Models}

The experiments for
$^{3}$He in aerogel are directly related to the corresponding measurements in pure $^{3}$He thereby
determining the effect of elastic scattering on specific combinations of $\beta$-parameters at
various pressures.  From the scattering models it is then possible to calculate the
model parameters, notably the transport mean free path, $\lambda$, and the impurity particle-particle
correlation length
$\xi_{a}$.  In the following we briefly review these scattering models and then discuss
experiments performed in 98\% aerogel and their  analysis in terms
of these parameters.  Despite possible variations between aerogel samples of the same nominal
porosity (see Fig.~8 as an example) we find that the values 
$\lambda = 150$ nm and
$\xi_{a} = 40$ nm fairly represent all of the experiments for 98\% aerogels and will be
discussed in detail in the following.

\subsection{Homogeneous Isotropic Scattering Model}

A number of theoretical models have been proposed to account for depairing of superfluid $^{3}$He
in aerogel.  The first results from Thuneberg {\it et al.}\cite{Thu96} and their subsequent
publication\cite{Thu98} provided a detailed framework that extended the Ginzburg-Landau theory for
impurities in superconductors\cite{Abr61} to
$p$-wave superfluidity.  The two important assumptions are
that i) the impurity scattering is homogeneous, and that ii) scattering is isotropic.  Consequently,
this model is called the homogeneous isotropic scattering model (HISM).  The HISM of  Thuneberg {\it
et al.}\cite{Thu98} modifies the $\beta$-parameters to accomodate depairing expressed in terms of
the AG-depairing parameter $x= \xi_{0}/\lambda$. Theoretical interpretation of experiments are
mostly based on this model or its
variations.\cite{Bar96,Thu98b,Sha01,Bar02,Min02,Han03,Sau03,Sha03,Sau05,Cho07} The modified
$\beta$-parameters, $\beta_{i}^a$ for the case of superfluid $^3$He in aerogel can be expressed as,

\begin{eqnarray} &\left(
\begin{array}{c}
\beta_{1}^{a}\\
\beta_{2}^{a}\\
\beta_{3}^{a}\\
\beta_{4}^{a}\\
\beta_{5}^{a}\\
\end{array}
\right) = \beta_{0}^{a}
\left(
\begin{array}{c} -1\\ 2\\ 2\\ 2\\ -2\\
\end{array}
\right) +b
\left(
\begin{array}{c} 0\\ 1\\ 0\\ 1\\ -1\\
\end{array}
\right) +
\left(
\begin{array}{c}
\Delta \beta_{1}^{sc,a}\\
\Delta \beta_{2}^{sc,a}\\
\Delta \beta_{3}^{sc,a}\\
\Delta \beta_{4}^{sc,a}\\
\Delta \beta_{5}^{sc,a}\\
\end{array}
\right),\\ &\beta_{0}^{a}={N(0) \over {30 (\pi k_{B}T_{c})^{2}}}
\sum_{n=1} {1 \over {(2n-1+x)^{3}}},\\ &b={N(0)\over{9(\pi k_{B}T_{c})^{2}}} \left( \mathrm{sin^{2}}
\delta_{0} - {1 \over 2}\right)
\sum_{n=1} {x \over {(2n-1+x)^{4}}}.
\end{eqnarray} 
The effects of scattering within the weak-coupling
approximation are included in $\beta_{0}^{a}$ and $b$; $\delta_{0}$ is the
$s$-wave scattering phase shift.
The strong-coupling corrections to the impurity-superfluid $\beta$-parameters,
$\Delta\beta^{sc,a}$, must be determined from experiment.   We take the point of view that
strong-coupling is manifest in the impurity system in the same way as for the pure superfluid with
the exception that these contributions are rescaled by the factor $T_{ca}/T_{c}$ allowing for the 
fact that strong-coupling in leading order is proportional to the transition
temperature.\cite{Rai76}  Then the prescription is straightforward.  The strong-coupling
contributions to $\Delta\beta^{sc}$ for the pure system are multiplied by the factor
$T_{ca}/T_{c}$ and added to the corresponding weak-coupling contributions for
the impure system, $\Delta\beta^{wc,a}$, calculated for the model scattering parameters,
$\lambda$, $\xi_{a}$, and
$\delta_{0}$, with
$T_{ca}$ determined self-consistently from,
\begin{equation}
\text{ln}{T_{c}  \over T_{c,a}} =  2 \sum_{n=1} \left ({1 \over {2n-1}} - {1
\over {2n-1+x}}\right).
\end{equation}
Values for all of the  
$\beta_{i}$'s have been discussed by Choi {\it et al.}\cite{Cho07}  Within the context of the
various scattering  models one can then estimate  stability of various superfluid states and their
properties. 

\begin{figure}[t]
%%%%%%%%%%%%%%%%%   F I G U R E  9   %%%%%%%%%%%%%%%%%%
\centerline{\includegraphics[width=2.8in]{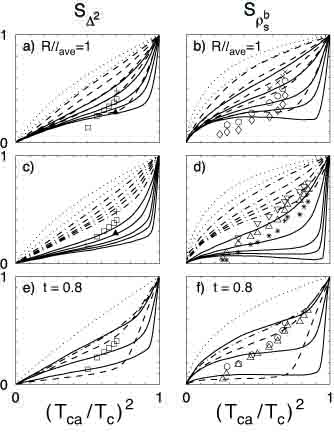}}
%%%%%%%%%%%%%%%%%%%%%%%%%%%%%%%%%%%%%%%%%%%%%
\begin{minipage}{0.93\hsize}
\caption {Suppression factors for the order parameter and the superfluid density calculated by
H\"anninen and Thuneberg\cite{Han03} for various scattering parameters in their theory and
at different temperatures. The solid curves are the IISM calculations with temperatures $t=0.2,
0.5,0.7,0.8, 0.9$ from top to bottom in a) to d).  The dot-dashed curves are the HISM results. For a)
and b), $R= \lambda$; c) and d), $R= 2\lambda$; and e) and f) top to bottom,
$R/\lambda=0.5, 1, 2$.  The data are from NMR,\cite{Spr95,Spr96} left side, and torsional oscillator
experiments,\cite{All98,Por99} right side.  This version of the IISM indicates less suppression of
the transition temperature as compared with the order parameter, that there is little distinction
between $A$ and
$B$-phases, and that the average superfluid density is more sensitive to inhomogeneity than the
order parameter amplitude. }
\end{minipage}
\end{figure}
\noindent

\subsection{Inhomogeneous Isotropic Scattering Models}

Several modifications of the HISM have been developed to include inhomogeneous scattering as well
as anisotropic scattering.\cite{Thu98,Han03,Sau03}  The most complete study of inhomogeneous
isotropic scattering (IISM) is that of H\"anninen and Thuneberg.\cite{Han03} In this
approach the $^3$He is contained in a spherical volume of radius $R$ within which there is a
spherically symmetric distribution of scattering centers.  The nature of the distribution is
defined by parameters in the theory.  The radius $R$ is chosen to be close to the
average transport mean free path, $\lambda$.  The  model is numerically intensive but, for suitable
parameter values as shown in Fig.~9, can lead to a fair comparison with experiment for both
superfluid density and the amplitude of the order parameter.  The main feature is that the
suppression of the order parameter and the superfluid density is greater than the suppression of
the transition temperature for the IISM, in agreement with experiment, and greater than the HISM,
dashed and dotted lines in Fig.~9.  One prediction of the IISM that is inconsistent with experiment
is the significant temperature dependence to the suppression factors. The NMR
experiments\cite{Spr95,Bau04} find that the measured frequency shifts scale precisely with the values
for the pure superfluid over a wide range of temperature, with a temperature independent
suppression factor.

Sauls and Sharma\cite{Sau03} have proposed a simple
phenomenological IIS-model that simply redefines the depairing parameter of Abrikosov and Gorkov
and is more immediately accessible for data analysis in the context of the Ginzburg-Landau theory.
They propose the modified depairing parameter $x=\hat x/(1+\zeta_{a}^{2}/\hat x)$,
$\zeta_{a}=\xi_{a}/\lambda$, where 
$\hat x=\hbar v_{F}/2 \pi k_{B}T \lambda$ is the depairing parameter in the HISM,  and
$\xi_{a}$ is  the aerogel particle-particle correlation length.  This form preserves the correct
limit for homogeneous scattering as $\xi_{a}$ goes to zero and in the opposite limit approaches
the form expected for randomly distributed voids of size $\xi_{a}$.  The effectiveness of this
model can be seen in Fig.~1.  The excellent agreement of the theory, solid blue curve (IISM), with
the data for transition temperatures is in contrast with the HISM, blue dashed curve.  It
appears that both IISM theories can successfully correct the difficulty with the HISM noted
above and shown in Fig.~8.  However, we will follow the approach suggested by  Sauls and Sharma owing
to its accessibility for data interpretation.

%%%%%%%%%%%%%%%%%%%%% TABLE 1 %%%%%%%%%%%%%%%%%%%%
\begin{table*}[t] {\small
\begin{tabular*}{0.8\textwidth}{@{\extracolsep{\fill}}c||ccccc|ccccc} & \multicolumn{5}{c|}{Pure
$^{3}$He} &\multicolumn{5}{c}{$^{3}$He in 98\% aerogel}\\*[2pt]
$P$ & $\frac{\beta_1}{\beta_0}$ & $ \frac{\beta_2}{\beta_0} $ & $\frac{\beta_3}{\beta_0}$ &
$\frac{\beta_4}{\beta_0}$ & $\frac{\beta_5}{\beta_0}$ & $\frac{\beta_1^{a}}{\beta_0^{a}}$ &
$\frac{\beta_2^{a}}{\beta_0^{a}}$ & $\frac{\beta_3^{a}}{\beta_0^{a}}$ &
$\frac{\beta_4^{a}}{\beta_0^{a}}$ & $\frac{\beta_5^{a}}{\beta_0^{a}}$
\\*[5pt] \hline\hline w.c. & -1 & 2 & 2 & 2 & -2 & -1 & 2 & 2 & 2 & -2
\\ \hline
 0 & -0.97 & 1.89 & 2.10 & 1.85 & -1.84 &  &  &  &  & \\*[-3pt]
 1 & -0.97 & 1.94 & 1.96 &1.72 & -1.82 &  &  &  &  & \\*[-3pt] 
 2 & -0.97 & 1.96 & 1.86 & 1.63 & -1.82 &  &  &  &  & \\*[-3pt]
 3 & -0.98 & 1.99 & 1.81 & 1.56 & -1.82 &  &  &  &  & \\*[-3pt]
 4 & -0.98 & 1.99 & 1.76 & 1.52 &-1.82 &  &  &  &  & \\*[-3pt]
 5 & -0.98 & 1.99 & 1.74 & 1.48 & -1.81 & -0.05 & 0.15 & 0.10 & 0.15 & -0.15\\*[-3pt]
 6 & -0.98 & 1.99 & 1.72 & 1.46 & -1.81 & -0.20 & 0.51 & 0.36 & 0.48 & -0.50\\*[-3pt] 
 7 & -0.98 & 1.98 & 1.70 & 1.44 & -1.81 & -0.28 & 0.72 & 0.53 & 0.66 & -0.70\\*[-3pt] 
 8 & -0.98 & 1.98 & 1.70 & 1.42 & -1.81 & -0.35 & 0.87 & 0.66 & 0.78 & -0.84\\*[-3pt] 
 9 & -0.99 & 1.98 & 1.69 & 1.41 & -1.81 & -0.41 & 0.99 & 0.75 & 0.87 & -0.95\\*[-3pt]  10 & -0.99 &
1.97 & 1.69 & 1.40 & -1.85 & -0.45 & 1.08 & 0.83 & 0.94 & -1.05\\*[-3pt]  11 & -0.99 & 1.97 & 1.70 &
1.39 & -1.85 & -0.49 & 1.15 & 0.90 & 0.99 & -1.12\\*[-3pt]  12 & -0.99 & 1.96 & 1.69 & 1.39 & -1.85
& -0.52 & 1.21 & 0.96 & 1.03 & -1.17\\*[-3pt]  13 & -0.99 & 1.95 & 1.69 & 1.39 & -1.85 & -0.55 &
1.26 & 1.01 & 1.06 & -1.22\\*[-3pt]  14 & -1.00 & 1.95 & 1.70 & 1.38 & -1.85 & -0.58 & 1.30 & 1.05 &
1.09 & -1.27\\*[-3pt]  15 & -1.00 & 1.95 & 1.72 & 1.35 & -1.89 & -0.60 & 1.34 & 1.10 & 1.10 &
-1.32\\*[-3pt]  16 & -1.00 & 1.95 & 1.73 & 1.34 & -1.89 & -0.62 & 1.38 & 1.13 & 1.12 &
-1.35\\*[-3pt]  17 & -1.00 & 1.94 & 1.72 & 1.33 & -1.89 & -0.64 & 1.40 & 1.16 & 1.13 &
-1.38\\*[-3pt]  18 & -1.00 & 1.94 & 1.73 & 1.32 & -1.89 & -0.66 & 1.43 & 1.19 & 1.14 &
-1.41\\*[-3pt]  19 & -1.00 & 1.93 & 1.72 & 1.33 & -1.89 & -0.67 & 1.45 & 1.21 & 1.16 &
-1.43\\*[-3pt]  20 & -1.01 & 1.94 & 1.74 & 1.31 & -1.93 & -0.69 & 1.48 & 1.24 & 1.16 &
-1.47\\*[-3pt]  21 & -1.01 & 1.94 & 1.74 & 1.29 & -1.93 & -0.71 & 1.50 & 1.26 & 1.16 &
-1.49\\*[-3pt]  22 & -1.01 & 1.93 & 1.74 & 1.29 & -1.93 & -0.72 & 1.51 & 1.28 & 1.17 &
-1.51\\*[-3pt]  23 & -1.01 & 1.93 & 1.74 & 1.29 & -1.93 & -0.73 & 1.53 & 1.30 & 1.18 &
-1.53\\*[-3pt]  24 & -1.01 & 1.93 & 1.74 & 1.28 & -1.93 & -0.74 & 1.54 & 1.32 & 1.18 &
-1.54\\*[-3pt]  25 & -1.01 & 1.93 & 1.74 & 1.28 & -1.97 & -0.75 & 1.56 & 1.33 & 1.18 &
-1.58\\*[-3pt]  26 & -1.02 & 1.93 & 1.73 & 1.27 & -1.97 & -0.76 & 1.57 & 1.34 & 1.18 &
-1.60\\*[-3pt]  27 & -1.02 & 1.93 & 1.74 & 1.26 & -1.97 & -0.77 & 1.58 & 1.36 & 1.18 &
-1.61\\*[-3pt]  28 & -1.02 & 1.93 & 1.73 & 1.26 & -1.97 & -0.78 & 1.60 & 1.37 & 1.19 &
-1.62\\*[-3pt]  29 & -1.02 & 1.93 & 1.73 & 1.26 & -1.97 & -0.78 & 1.61 & 1.38 & 1.19 &
-1.63\\*[-3pt]  30 & -1.02 & 1.93 & 1.72 & 1.26 & -2.01 & -0.79 & 1.62 & 1.38 & 1.19 &
-1.67\\*[-3pt]  31 & -1.03 & 1.93 & 1.73 & 1.25 & -2.01 & -0.80 & 1.62 & 1.40 & 1.19 &
-1.68\\*[-3pt]  32 & -1.03 & 1.93 & 1.73 & 1.25 & -2.01 & -0.81 & 1.63 & 1.40 & 1.19 &
-1.68\\*[-3pt]  33 & -1.03 & 1.93 & 1.73 & 1.25 & -2.01 & -0.81 & 1.63 & 1.41 & 1.19 &
-1.69\\*[-3pt]  34 & -1.03 & 1.93 & 1.73 & 1.25 & -2.01 & -0.82 & 1.64 & 1.42 & 1.20 & -1.70
\end{tabular*} }
\caption{$\beta_{i}$'s for pure superfluid $^{3}$He, left side, and  superfluid $^{3}$He in aerogel
in IISM with $\lambda$= 150 nm and $\xi_{a}$=40 nm, right side. These results are taken from Choi
{\it et al.}\cite{Cho07}.}
\label{beta}
\end{table*}
%%%%%%%%%%%%%%%%%%%%%%%%%%%%%%%%%%%%%%%%%%%%%%

\section{Experimental Results}

\subsection{Identification of the Superfluid State and Phase Diagram}

The first NMR experiments\cite{Spr95,Spr96,All98,Bar00} showed that paramagnetic solid $^3$He,
adsorbed on the aerogel surface, dominated the temperature dependent magnetization and consequently
reduced the NMR frequency shifts substantially by the ratio of the liquid to the total
magnetization  according to a model for fast exchange.  These effects can be allowed for in the
measurement as is the case for NMR data in Fig.~7, or they can be eliminated by precoating the
surfaces\cite{Spr96} with two or more atomic layers of
$^4$He.  It seems that  preplating with
$^4$He has only slight effect on the overall phase diagram.\cite{Spr96}

Identification of the symmetry of the observed phases was not possible from
measurements of the superfluid density with the torsional oscillator technique.\cite{Por95}  On
the other hand early experiments on NMR frequency shifts and spectra were interpreted either as an
equal-spin-pairing (ESP) state,\cite{Spr95} like the
$A$-phase with a uniform texture (dipole-locked); or the $B$-phase with an inhomogeneous
texture\cite{All98} determined by sample shape (dipole-unlocked).  It seems likely that these very
different results and their interpretation are a consequence of  metastability of the
$A$-like phase and  differences between the two samples. 

A clear signature of an ESP-state is its temperature
independent magnetization found from the integral of the NMR spectrum.  In contrast, the
$B$-phase magnetization is temperature dependent and a transition
from an
$A$-like phase to a $B$-like phase is marked by a discontinuous change in the magnetization.  This
was first seen by Barker
{\it et al.}\cite{Bar00} in a 98\%  aerogel and subsequently by many other
groups.\cite{Dmi03,Bau04,Bun05,Nak05,Kun07,Nak07} Additionally the frequency spectrum in most NMR
experiments is substantially  broadened inhomogeneously owing  to a distribution in texture of the
order parameter whose orientation relative to the magnetic field can generate a wide range of
frequency shifts, $\Delta\omega$, which can be expressed\cite{Vol90} for pure $^3$He as, 

\begin{equation}
\omega^2 = \omega^2_{0} + F_{A,B}(\theta,\phi)\Omega^2_{A,B}
\end{equation}
\begin{equation}
\Delta\omega \equiv \omega_{0} - \omega \simeq
F_{A,B}(\theta,\phi) \frac{\Omega^{2}_{A,B}}{2\omega_{0}},
\end{equation}

\begin{figure}[b]
%%%%%%%%%%%%%%%%%   F I G U R E  10   %%%%%%%%%%%%%%%%%%
\centerline{\includegraphics[width=2.8in]{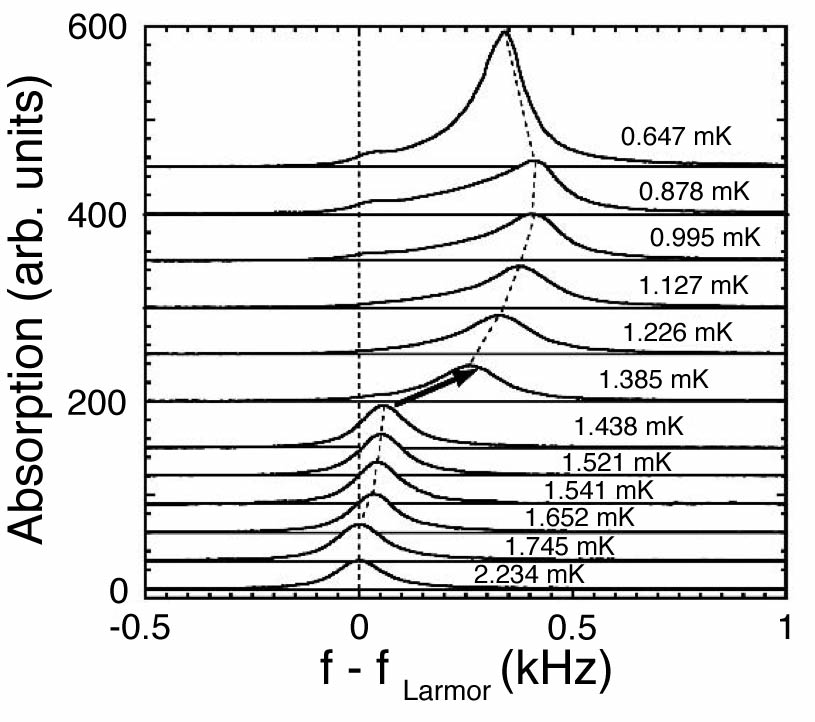}}
%%%%%%%%%%%%%%%%%%%%%%%%%%%%%%%%%%%%%%%%%%%%%
\begin{minipage}{0.93\hsize}
\caption {Temperature dependence of the NMR spectra passing from the normal to $A$-like phase and
then to the $B$-like phase for a 97.5\% aerogel, uncorrected for surface solid $^3$He, obtained by
Nakagawa {\it et al.}\cite{Nak05} If maximum shifts in the NMR spectra, allowing for the normal
state lineshape,  are identified with the longitudinal resonance frequency,
$\Omega_{A,B}$, {\it i.e.} $F_{A}=1, F_{B}=0.8$ in Eq.~18, then the relative magnitude of the shifts
in the spectra appear to be roughly consistent with a transition from the axial state to the isotropic
state, Eq.~20, provided the angular momentum vector is perpendicular to the field, $\ell \perp H$.
(See the discussion by Elbs {\it et al.}\cite{Elb08})}
\end{minipage}
\end{figure}
\noindent
where $\Omega_{A,B}$ is the longitudinal resonance frequency for the $A$ or $B$-phase, $\omega_{0}$
is the NMR Larmor frequency,  and
$F_{A,B}(\theta,\phi) \leq 1$ depends on the NMR tip angle, $\phi$, and the orientation of the order
parameter, $\theta$. However, the existence of negative shifts for $\phi \approx 0$ (the cw-NMR case
for example,) even if only in some portions of the spectrum, is a direct indication\cite{Bar00} of
a phase other than
 the isotropic phase.  The isotropic phase can only have positive shifts; so negative shifts can be
interpreted as an indication of an $A$-like phase, a point that was emphasized by Barker
{\it et al.}\cite{Bar00}  For the axial($A$) and isotropic($B$) states we know further
that,\cite{Leg75}
 \begin{equation}
\frac{\Omega_{B}^2}{\Omega_{A}^2} = \frac{5}{2}
\left(\frac{\chi_{A}}{\chi_{B}}\right) \frac{\Delta_{B}^2}{\Delta_{A}^2}.
\end{equation}
If we identify $B$ and $A$-like phases, respectively with
the isotropic and axial states, then close to $T_{ca}$ where
$\Delta_{A} \approx \Delta_{B}$ (since $\beta_{A} \approx \beta_{B}$ from table I) and $\chi_{A} =
\chi_{B}$, we expect the largest frequency shifts of the
$B$-like phase to be two times larger than for the $A$-like phase.  The ratio of frequency shifts
observed for the two phases in a 97.5\% aerogel by Nakagawa {\it et al.},\cite{Nak05} reproduced in
Fig.~10, appear to be consistent with this behavior provided $\ell \perp H$. Baumgardner and
Osheroff\cite{Bau04b} have made similar arguments to determine the free energy of $A$-like and
$B$-phases in a 99.3\% aerogel, and Dmitriev {\it et al.}\cite{Dmi08} have also analyzed specta for a
98\% aerogel to obtain the relative magnitudes of the longitudinal resonance frequencies in the two
phases.  In both cases the spectra are complex owing to a distribution of textures and, in the latter
case, coexistence of these two phases in different parts of their aerogel sample. Direct observation
of the longitudinal resonance frequency has also been reported by this group in both
phases\cite{Dmi06} as well as observation\cite{Dmi03b} of a homogeneous precession domain (HPD) in
the aerogel $B$-phase.  Sato {\it et
al.}\cite{Sat08} have observed HPD in an ESP-state of superfluid
$^3$He for the first time using aerogel.  This was obtained in the $A$-like phase in a sample
that was axially anisotropic with symmetry axis parallel to the magnetic field.  This arrangement
has  maximum dipole energy and leads to maximal negative shifts in the spectrum; however, the tip
angle dependence of the dipole energy provides a favorable situation for the formation of a HPD in
this case, not realized in pure $^3$He.

\begin{figure}[t]
%%%%%%%%%%%%%%%%%   F I G U R E  11   %%%%%%%%%%%%%%%%%%
\centerline{\includegraphics[width=2.8in]{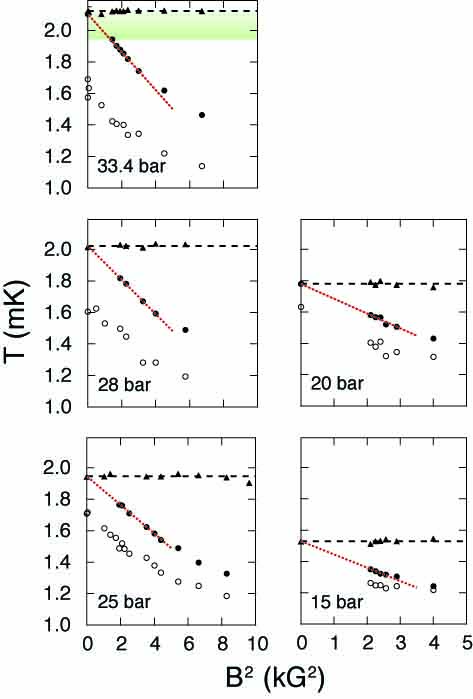}}
%%%%%%%%%%%%%%%%%%%%%%%%%%%%%%%%%%%%%%%%%%%%%
\begin{minipage}{0.93\hsize}
\caption {Phase diagram for a 98\% aerogel superfluid in a magnetic field from Gervais {\it et
al.}\cite{Ger02}  An equal-spin pairing-state like the $A$-phase was identified from the independence
of its transition temperature on field.  This phase destabilizes the dominant
$B$-phase with increasing magnetic field at
all pressures studied.  The temperature interval for metastability of the $A$-like
phase (open circles) appears to be independent of field, on which basis Gervais {\it et al.} inferred
that the window of stable $A$-like phase in zero magnetic field was very small.  This argument is
based on the assumption that the $B$ to $A$-like phase transition does not superheat, as is the case
for pure $^3$He, and the fact that the zero-field extrapolation of the warming transition points
directly toward
$T_{ca}$, as is shown in this figure.}
\end{minipage}
\end{figure}
\noindent

\begin{figure}[h]
%%%%%%%%%%%%%%%%%   F I G U R E  12   %%%%%%%%%%%%%%%%%%
\centerline{\includegraphics[width=2.8in]{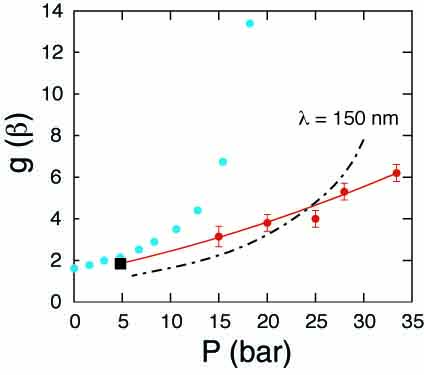}}
%%%%%%%%%%%%%%%%%%%%%%%%%%%%%%%%%%%%%%%%%%%%%
\begin{minipage}{0.93\hsize}
\caption {The quadratic field suppression of the $B$-phase in a 98\% aerogel from Gervais {\it
et al.}\cite{Ger02}  These data are taken from the phase diagram measurements in Fig.~11. The same
factor
$g(\beta)$ was obtained for either warming or cooling and is shown here compared to their fit
(dot-dashed curve) to the HISM giving $\lambda=150$ nm. The black square is taken from Brussard {\it
et al.}\cite{Bru01} Pure $^3$He is shown by solid (blue) circles\cite{Tan91} and the solid (red) curve
is a guide to the eye.}
\end{minipage}
\end{figure}
\noindent

\begin{figure}[t]
%%%%%%%%%%%%%%%%%   F I G U R E  13   %%%%%%%%%%%%%%%%%%
\centerline{\includegraphics[width=2.8in]{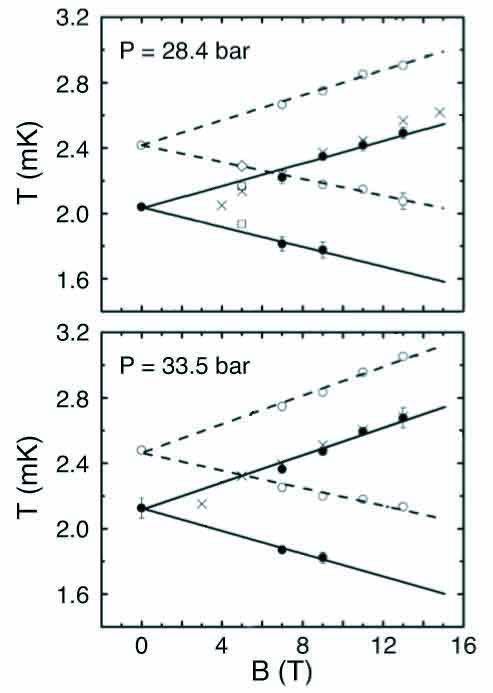}}
%%%%%%%%%%%%%%%%%%%%%%%%%%%%%%%%%%%%%%%%%%%%%
\begin{minipage}{0.93\hsize}
\caption {Splitting of the $A_{1} - A_{2}$ transition at high magnetic field for superfluid $^3$He
 in a 98\% aerogel (solid circles and crosses) and in pure $^3$He (open circles) at two pressures by
Choi {\it et al.}\cite{Cho04hc}  The linear slopes with magnetic field are rather similar for pure
and aerogel superfluids.}
\end{minipage}
\end{figure}
\noindent

Complementary to the NMR methods, which require an applied field,
acoustic measurements in zero applied field\cite{Ger01,Ger02,Naz04} can identify phase transitions
with excellent resolution.  Then one can determine which of those transitions are associated with
$A$-like or
$B$-like phases by applying a field, the isotropic phase being strongly suppressed by field as
compared to an ESP-phase. Based on such methods Gervais {\it et al.}\cite{Ger02} concluded that the
entire pressure-temperature, superfluid region in zero applied magnetic field for a 98\% aerogel was
a $B$-like phase, with the possible exception of a narrow region just below
$T_{ca}$, $\approx 20 \mu$K wide.  In this $B$-like phase Dmitriev {\it et al.}\cite{Dmi02} determined
a Leggett angle equal to $104^\circ$ from pulsed NMR measurements as a function of pulse tip-angle,
thereby confirming that the
$B$-like phase was the isotropic state similar to pure superfluid $^3$He.  In the following we will
refer to this phase simply as the $B$-phase.     

Using transverse acoustic impedance, Gervais {\it et al.}\cite{Ger02} measured the
magnetic field suppression of the $B$-phase for several representative
pressures from 15 to 33.4 bar, Fig.~11.    On cooling in a magnetic field the metastable
$A$-like phase was first nucleated at a field-independent value, $T_{ca}$, in the range from
zero-field to 0.5 T.  The field dependence for transitions from $A$-like to $B$-phases on cooling was
quadratic in field.  On warming the  $B$ to $A$-like transition was displaced to a higher temperature
but with the same quadratic field dependence.  It is likely  that the magnetic field dependence of
the equilibrium transition is the same as found in these warming and cooling experiments since
those data are parallel to one another.  Gervais {\it et al.} determined
$g(\beta)$ from their analysis of this field dependence in terms of Eq.~11 and 12.  The HISM can account for
the observed behavior with a scattering mean free path of $150$ nm, Fig.~12. Brussard {\it et
al.}\cite{Bru01}  determined the field dependence of the $A$-like to $B$-transition at
4.8 bar with an oscillating aerogel paddle and these results are consistent, shown by the black
square in Fig.~12.

The $A$-like phase is known to be an ESP-phase even if it is not well-established to which $p$-wave
state this corresponds.  For sufficiently large magnetic fields an ESP-state should split into
$A_{1}$-like and $A_{2}$-like transitions owing to particle-hole asymmetry.  Barmadze and
Kharadze\cite{Barm00} and Sauls and Sharma\cite{Sau03} have considered this problem for superfluid
$^3$He in aerogel taking into account the exchange coupling between the paramagnetic surface solid
$^3$He and the superfluid which will either decrease or increase the splitting depending on whether
the interaction is antiferromagnetic or ferromagnetic.  Since there did not appear to be evidence
for a splitting in experiments\cite{Ger02} up to 0.8 T this suggested that the coupling was
antiferromagnetic. But higher field experiments are required, as well as measurements with
$^4$He preplating, to reach a firm conclusion. In this spirit Choi {\it et al.}\cite{Cho04hc}
extended the field range of Gervais {\it et al.}\cite{Ger02} to $H=14$ T using the same aerogel
sample, and their results are reproduced in Fig.~13. Transverse sound techniques were used and the
$A_{1}$-$A_{2}$ splitting was clearly identified in both pure and aerogel superfluids.  The
splitting ratio,
$r$, was found to be similar for the two cases. Owing to experimental conditions, the accuracy
of the transition temperatures measured was not sufficient at low magnetic fields to make a
quantitative comparison with the exchange models.\cite{Barm00,Sau03} 

\begin{figure}[t]
%%%%%%%%%%%%%%%%%   F I G U R E  14   %%%%%%%%%%%%%%%%%%
\centerline{\includegraphics[width=2.8in]{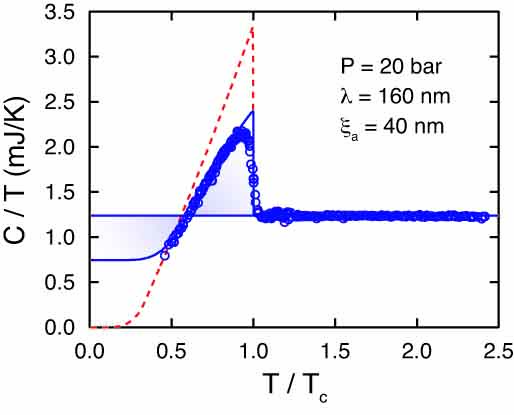}}
%%%%%%%%%%%%%%%%%%%%%%%%%%%%%%%%%%%%%%%%%%%%%
\begin{minipage}{0.93\hsize}
\caption {Specific heat jump as a function of temperature for the $B$-phase of a 98\% aerogel at
$P = 20$ bar.  Fitting to the data gives the scattering parameters  of
mean-free-path $\lambda = 160$ nm and $\xi_{a} = 35$ nm from the IISM of Sauls and
Sharma.\cite{Sau03}  The specific heat for pure $^3$He is shown for comparison as a dashed
curve.\cite{Gre86}}
\end{minipage}
\end{figure}
\noindent

\begin{figure}[b]
%%%%%%%%%%%%%%%%%   F I G U R E  15   %%%%%%%%%%%%%%%%%%
\centerline{\includegraphics[width=2.8in]{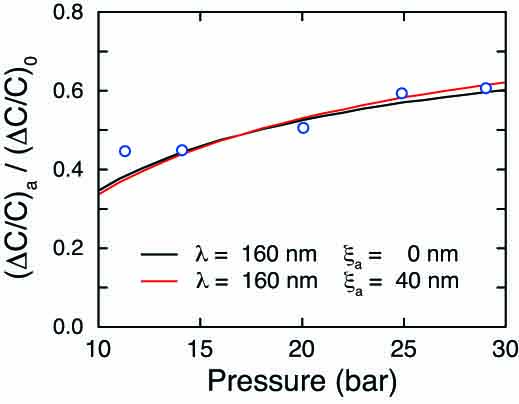}}
%%%%%%%%%%%%%%%%%%%%%%%%%%%%%%%%%%%%%%%%%%%%%
\begin{minipage}{0.93\hsize}
\caption {Specific heat jump as a function of pressure for the $B$-phase of a 98\% aerogel
from Choi {\it et al.}\cite{Cho04}  Fitting to the data gives the scattering parameters  for
the IISM of Sauls and
Sharma\cite{Sau03} of mean-free-path $\lambda = 160$ nm and $\xi_{a} = 40$ nm. The fits are
not sensitive to the correlation length, $\xi_{a}$ as
shown here, in contrast to the transition temperature, Fig.~1 and 8. Additionally, there is only a
small difference between $A$- and $B$-phases although this is not shown here.\cite{Cho07}}
\end{minipage}
\end{figure}
\noindent

\subsection{Specific Heat, Thermal Conductivity, Gaplessness}

The most direct determination of the magntitude of the order parameter is found from the specific
heat jump, a measurement that can be interpreted in terms of the GL-theory and which determines
corresponding values of the $\beta$-parameters.  Specific heat measurements were performed by Choi
{\it et al.}\cite{Cho04} as shown in Fig.~14 and 15.  In this experiment an adiabatic calorimetric
method was used, hampered in part by the small volume of the aerogel sample, 1.1 cm$^3$, 
and a
background  attributed to paramagnetic solid on the aerogel surface that was subtracted.  On the
other hand, very low heat leak, $\approx 0.1$ nW,  and high resolution thermometry gave accurate
measurements, particularly for the specific heat jump.  A comparison of the jump for pure and
aerogel superfluids is made in Fig.~15.  Good agreement was found with the predictions of the theory
as a function of pressure, leading to a best fit overall for $\lambda = 160$ nm.  The
distinction between HISM and IISM theories is marginal in this case.  Additional
calculations\cite{Cho07thesis} for the specific heat jump within the context of these models for
$A$ and $B$-phases indicate that they are close to one another ($5$ to
$10$\%, depending on pressure) although
$\Delta C_{Ba}$ is always larger than $\Delta C_{Aa}$, consistent with the equilibrium phase
being the $B$-phase. A comparison of the order parameter suppression and that of the
superfluid transition temperature for this 98\% aerogel is plotted in Fig.~8.

\begin{figure}[t]
%%%%%%%%%%%%%%%%%   F I G U R E  16   %%%%%%%%%%%%%%%%%%
\centerline{\includegraphics[width=2.8in]{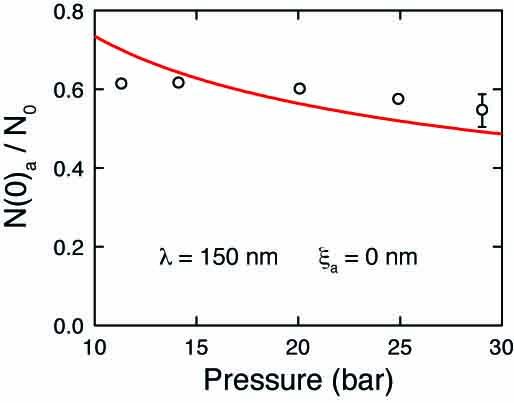}}
%%%%%%%%%%%%%%%%%%%%%%%%%%%%%%%%%%%%%%%%%%%%%
\begin{minipage}{0.93\hsize}
\caption {The density of states (DOS) at the Fermi energy inferred from specific heat
measurements of Choi {\it et al.}\cite{Cho04}   compared with calculations from the HISM.  The
non-zero spectral weight at the Fermi energy is attributed to overlapping bound states.  This
behavior is a direct indication of gapless superfluidity and has been observed and predicted for
the thermal conductivity,\cite{Sha03,Fis03} and the magnetic susceptibility.\cite{Min02, Bar00}}
\end{minipage}
\end{figure}
\noindent

In addition to the specific heat jump, Choi {\it et al.}\cite{Cho04}  determined the temperature
dependence of the specific heat and in particular accurate measurements of the temperature derivative
of the specific heat just below $T_{ca}$.  They argued on the basis of the third law of
thermodynamics that entropy should be conserved, meaning that the two shaded areas in Fig.~14 must
be equal and consequently if the specific heat is a monotonic function of temperature that there is
a non-zero intercept of $C_{Ba}/T$  at $T=0$.  This observation means that
there is a non-zero and significant density of states (DOS) at the Fermi energy, $E_{F} \equiv 0$. 
This behavior is in contrast to pure superfluid  $^3$He where the DOS is strictly zero there. 
Theoretically it is expected\cite{Sha01,Min02} that a broad band of Andreev bound states associated
with the aerogel impurity in $^3$He will transfer spectral weight in the DOS from the coherence
peaks above $\Delta$ to zero energy, leading to gapless behavior.  With reduction of
pressure the DOS band develops, eventually reaching a value close to that of the
normal state, in which case the specific heat on passing from the normal to the superfluid becomes
featureless. It is also clear that the development of this band of states obscures the distinction
between different possible
$p$-wave states.  Consequently, the presence of large concentrations of impurities in an
unconventional superfluid or superconductor will substantially affect the low
temperature thermodynamic behavior obscuring identification of the symmetry of the
superfluid or superconducting state in these measurements. The observation of non-zero values of the
DOS at low energy from the specific heat is plotted as a function of pressure in Fig.~16 and
compared with the calculations of Sharma and Sauls.\cite{Sha01} For these data they 
found\cite{Cho04} a best fit value of
$\lambda = 150$ nm.  Evidence for gapless behavior can also be found in the low temperature magnetic
susceptibility\cite{Min02,Sau05,Hal04} and the thermal conductivity.\cite{Fis03} In the latter case
the thermal conductance indicates that a gap in the $B$-phase DOS returns for pressures in excess of
about 10 bar, in contrast to the interpretation of gapless behavior at all pressures from the specific
heat measurements made by Choi {\it et al.}

Thermal conductivity measurements have been performed by the Lancaster group,\cite{Fis01,Fis03} for
which the high temperature superfluid behavior is presented in Fig.~17, showing that there is a weak
anomaly at the transition temperature.  Behavior in the normal fluid and in the superfluid phases has
been analyzed by Sharma and Sauls\cite{Sha03} from which 
$\lambda = 205$ nm was determined, somewhat larger than the trend from other experiments.  It is
possible that the thermal boundary resistance between the aerogel and pure $^3$He superfluids
might play a significant role.

\begin{figure}[h]
%%%%%%%%%%%%%%%%%   F I G U R E  17   %%%%%%%%%%%%%%%%%%
\centerline{\includegraphics[width=2.8in]{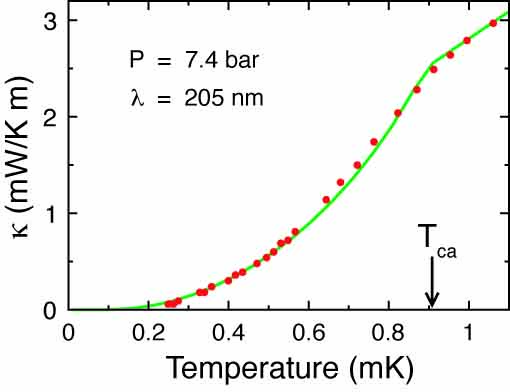}}
%%%%%%%%%%%%%%%%%%%%%%%%%%%%%%%%%%%%%%%%%%%%%
\begin{minipage}{0.93\hsize}
\caption {Thermal conductivity data at 7.4 bar from Fisher {\it et al.}\cite{Fis01}
analyzed\cite{Sha03} in terms of the HISM theory giving a value of $\lambda = 205$ nm.}
\end{minipage}
\end{figure}
\noindent

\subsection{NMR and Magnetization}

The magnetization measurements\cite{Spr96,Bar00,Bau04,Sau05,Bun05} consistently indicate a
metastable ESP-phase and a $B$-phase with a suppressed order parameter.  Calculations have been
performed by Sharma and Sauls\cite{Sha03} and Mineev\cite{Min02} for which a representative example
of data is shown in Fig.~18 from Barker {\it et al.}\cite{Bar00}  A value of
$\lambda = 140$ nm fits the data best and the distinction between
HISM and IISM is not discernible. These experiments were performed with $^4$He preplating.  

In the context of identification of the superfluid state we have discussed the NMR
frequency shifts and spectra expected for the $A$ and $B$-phases. In some cases, such as the 99.3\%
aerogel samples of Baumgardner {\it et al.}\cite{Bau04} the spectrum is very broad; in other
cases\cite{Spr95,Nak07,Elb08,Kun07} much less so. In the first NMR experiments, Sprague {\it et
al.}\cite{Spr95} took data only on warming.  It is not obvious how to reconcile their
measurements with their claim that the experiments were conducted in the (supercooled) $A$-like phase
since the extent of supercooling reported by other workers\cite{Ger01,Ger02,Naz04} is typically a
factor of two less than the range of supercooling that would be required by Sprague {\it et al.}   
On the other hand, they measured the magnetization with sufficient accuracy to identify an ESP-phase.
Later analysis by Haard,\cite{Haa01} based on independent measurements of the magnetization, 
confirmed the earlier conclusions. Additionally, Sprague {\it et al.} found very little inhomogeneous
broadening of their spectra, in contrast to what has been found by most groups  to be characteristic
of the $B$-phase.  Finally, the amplitude of the order parameter inferred from frequency shifts
measured by Sprague {\it et al.} are in excellent agreement with specific heat jump measurements as
displayed in Fig.~8 and 19.  If the data of Sprague {\it et al.} were indeed in the $B$-phase, the
texture of the order parameter would necessarily have to be dipole-unlocked, but homogeneously so, in
order to have a well-defined value of $F(\theta,\phi)$ in Eq.~18. This seems rather unlikely although
it cannot be ruled out.  In summary, we believe that Sprague {\it et al.} observed extensive
supercooling of the metastable
$A$-like phase in those early NMR experiments.  We have recently inspected the
aerogel sample they used with optical birefringence methods
and found that it was homogeneous and isotropic.\cite{Pol08} Furthermore, the aerogel
sample was well separated ($\approx 0.2$ mm) from the walls eliminating the possibility of
stress accumulation from differential contraction during cooling. These aspects may account for the
homogeneous NMR spectra and  large supercooling.

\begin{figure}[h]
%%%%%%%%%%%%%%%%%   F I G U R E  18   %%%%%%%%%%%%%%%%%%
\centerline{\includegraphics[width=2.8in]{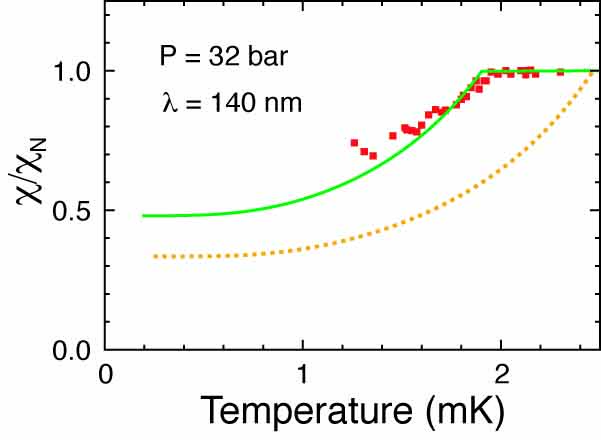}}
%%%%%%%%%%%%%%%%%%%%%%%%%%%%%%%%%%%%%%%%%%%%%
\begin{minipage}{0.93\hsize}
\caption {Magnetic susceptibility measurements in the $B$-phase of superfluid $^3$He in a 98\%
aerogel at
$P=32$ bar from NMR by Barker {\it et al.}\cite{Bar00} analyzed by Sharma and Sauls\cite{Sha01} in
terms of the HISM.  A value of
$\lambda = 140$ nm was found although the temperature range of the experiment is not extensive.  The
distinction between HISM and IISM is rather slight.  The susceptibility for pure
$^3$He is shown as a dotted curve. Similar work has been reported by the Grenoble group\cite{Sau05}
for a 98\% aerogel where they consistently found
$\lambda = 140$  for data in the range $P=17$ to 30 bar.}
\end{minipage}
\end{figure}
\noindent

\begin{figure}[b]
%%%%%%%%%%%%%%%%%   F I G U R E  19   %%%%%%%%%%%%%%%%%%
\centerline{\includegraphics[width=2.8in]{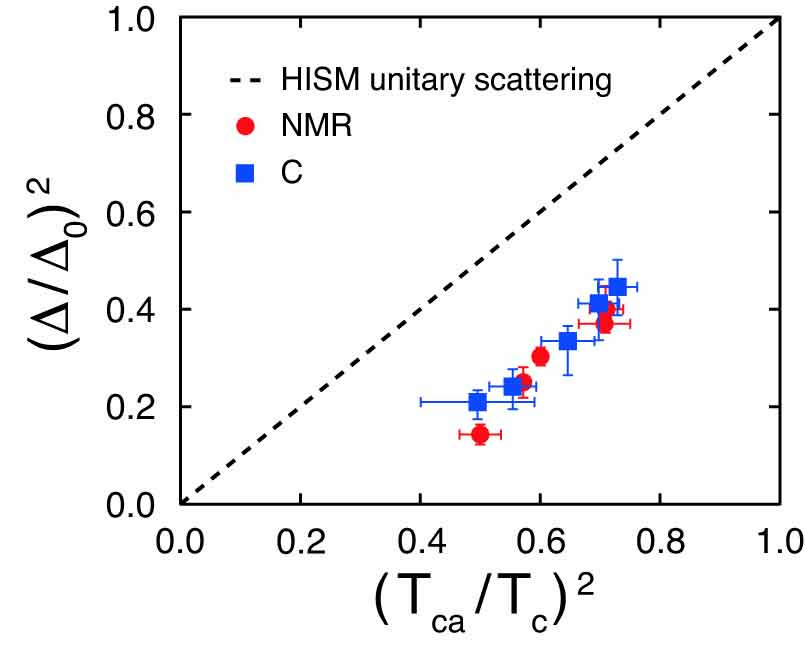}}
%%%%%%%%%%%%%%%%%%%%%%%%%%%%%%%%%%%%%%%%%%%%%
\begin{minipage}{0.93\hsize}
\caption {A direct comparison of the order parameter suppression relative to the transition
temperature is shown for 98\% aerogel, compared to the HISM with unitary scattering shown as a
dashed line. The excellent agreement between the specific heat data\cite{Cho04} and the frequency
shift measurements\cite{Spr95,Haa01} give support for the correctness of the phase
identification made by Sprague {\it et al.}\cite{Spr95}}
\end{minipage}
\end{figure}
\noindent

\subsection{Acoustics and Mechanical Response}

Low frequency
sound\cite{Gol99,Naz04} and high frequency ultrasound propagation have been
observed\cite{Nom00,Ger02,Cho07hc} in superfluid
$^3$He in aerogels and these techniques, combined with transverse acoustic impedance
methods,\cite{Ger01,Ger02} have been used to identify phase transitions as well as the superfluid
density. Calculations\cite{Hig05,Cho07hc} have focused on the
coupling between the silica framework and the $^3$He modes and yielded excellent agreement with
experiment including the collision drag effect. 

The mechanical dynamics of aerogel samples affixed to a wire
resonator have also been used by the Lancaster group\cite{Bru01, Bra07} to investigate the phase
diagram and have been exploited to determine the transitions between phases and anisotropy in the
superfluid density which may be relevant in distinguishing between various possible candidates for
the metastable $A$-like phase.  

At this time there is no evidence of a propagating high frequency
shear mode in superfluid $^3$He in aerogel and there is no evidence from acoustics of order parameter
collective modes\cite{Nom00,Cho07hc} nor an anomaly in attenuation associated with a threshold for
pair breaking.  These features have been extensively explored in the pure
superfluid phases.\cite{Hal90}

\subsection{ Summary of Comparison of the IISM with Experiment}

In the previous sections we have discussed applications of the scattering models to the analysis of 
experiments on superfluid $^3$He in aerogel.  These give a remarkably consistent picture. 
The scattering model parameters which we have found to be
typical of a 98\% porosity aerogel are:  $\lambda
= 150$ nm, $\xi_{a} = 40$ nm assuming unitary scattering, $\delta_{0}=\pi/2$.  These
parameters for the IISM can account for most experiments including the phase diagram, specific
heat, thermal conductivity, magnetic suscpetibility, superfluid density, NMR 
frequency shifts, and the high field $A_{1}$ - $A_{2}$ splitting of the normal to superfluid
transition.  With the five
$\beta_{i}$'s for pure superfluid given in Table I, we use the Sauls and Sharma IISM-approach  to
define a modified depairing parameter and  calculate the
$\beta_{i}$'s for the impure superfluid.   The results
of the calculation of
$\beta_{i}^{a}/\beta_{0}^{a}$ are tabulated in the last five columns of Table I.  For these 
parameters the superfluid state is not stable below a pressure of 5  bar as reported by
Matsumoto {\it et al.}\cite{Mat97}  A direct
consequence of the modification of $\beta_{i}$'s, according to the scattering models
is the enhancement of  relative stability of the
$B$-phase with respect to the $A$-phase over the entire  pressure range below the melting
pressure of
$^{3}$He. Consequently, the existence of a metastable anisotropic phase  suggests that the form of
the free energy in Eq.~2 is incomplete.

\section{Metastable $A$-like Phase}

Superfluid
$^{3}$He in aerogel has a metastable  phase that has
been  clearly observed\cite{Spr95,Bar00,Ger01,Ger02,Naz04,Bru01,Bra07,Bau04} in various samples
appearing on cooling below
$T_{ca}$ in a magnetic field.  It is natural to expect that a window of
$A$-phase might open up in a magnetic field and that this phase might supercool
substantially, possibly even at low pressures.  The exact nature of
this phase  remains an open question, although NMR experiments have shown that the magnetic
susceptibility is unchanged from the normal state indicating that it is an ESP-state, like the
$A$-phase.\cite{Spr95,Bar00} It is significant that Gervais {\it et al.}
showed\cite{Ger01}  that the same phenomenon of metastability of an $A$-like phase occurs in
experiments in {\it zero applied field}; evidently the mechanism for nucleation and the corresponding
metastability is independent of magnetic field. Additional work at various pressures in zero
applied field lead to the same conclusions.\cite{Ger02,Naz04} Based on these
experimental results and from the theoretical suggestions of Thuneberg {\it et al.}\cite{Thu98},
Vicente {\it et al.}\cite{Vic05} proposed that anisotropic scattering might be the origin for
stability of the $A$-like phase.  However, within the context
of the free energy expansion, Eq.~2, the question
of stability of any ESP-state with repect to  the aerogel $B$-phase  relies on an
understanding of the appropriate
$\beta$-parameters, at least in the Ginzburg-Landau limit close to $T_{ca}$. The relative stability
of axial and isotropic phases  can be determined for IISM from the $\beta$-parameter
combinations taken from Table I that give $\Delta C_{Aa}$ and $\Delta C_{Ba}$, Eq.~9 and 10.  The two
phases are thermodynamically so close that they will be difficult to distinguish directly
$(5\sim10\%)$ in specific heat measurements.\cite{Cho07thesis} Nonetheless, as we have mentioned
previously, it is likely that anisotropic scattering or some other modification of the theory will
be required to understand this interesting phenomenon.  

Volovik\cite{Vol96} has argued that the axial state in the
presence  of quenched anisotropic disorder  cannot exist as a spatially homogeneous  superfluid. 
So the axial state would not have long range orientational order, a state which Volovik has
called  a superfluid glass or a Larkin-Imry-Ma state.\cite{Imr75}  With a  different approach,
Fomin\cite{Fom04} has argued that there are other
$p$-wave pairing states which are also ESP-states but  do not suffer from the same
difficulty, and that these might be candidates for the  metastable aerogel phase.  Such phases
would be robust in the  presence of anisotropic scattering, meaning that
$A_{\mu i}A_{\mu  j}^{*} + A_{\mu j}A_{\mu i}^{*} \varpropto \delta_{ij}$ where $\delta_{ij}$ is a
Kronecker delta.\cite{Fom04} NMR experiments have been performed on
$^{3}$He in
$97.5\%$ aerogel  which support the view that the metastable $A$-like phase  is in fact a robust
state,\cite{Ish06}  but other measurements
\cite{Dmi06,Dmi08} appear to have inconsistencies with this interpretation.   

If we assume that the free energy expansion of Eq.~2 is correct then we can use the
$\beta_{i}$'s from Table I to  determine the asymmetry ratio predicted for the robust and
axial states and compare with experiment.\cite{Cho04hc}  For the  axial state  the ratio is expressed
in terms of the 
$\beta_{i}$'s given by Eq.~13 for which we find $r$ between 1.0 and 1.5, increasing
with increasing pressure consistent with the experimental results $r_{a} \gtrsim 1.0$.\cite{Cho04hc} 
In the case of the robust state\cite{Fom04,Cho04},
\begin{equation} r_{R} ={\beta_{15} \over {\beta_{12345}+4\beta_{2}+4\beta_{245}}}.
\end{equation} With the values of the $\beta_{i}$'s from the table, the asymmetry  ratio $r_{R}$ is
found to be $\sim 0.2$, considerably smaller than what was observed. 

\section{Anisotropic Aerogel}

There are two important issues regarding anisotropy.  First, according to
the scattering models\cite{Thu98} isotropic elastic scattering stabilizes the isotropic state and
anisotropic scattering favors an anisotropic state. Secondly, all experiments are consistent with the
formation of a metastable anisotropic phase (there is only one isotropic $p$-wave state) which
somehow nucleates on cooling through $T_{ca}$.  To reconcile these points Vicente {\it et
al.}\cite{Vic05} proposed that the local anisotropy inherent to the aerogel structure on the scale of
$\xi_{a}$  should lead to anisotropic scattering that could nucleate an anisotropic phase.
Furthermore this effect will be more prevalent for shorter coherence lengths, {\it i.e.} at higher
pressure.  The physical mechanism for this phenomenon is completely independent of that which
stabilizes the
$A$-phase in pure superfluid $^3$He {\it i.e.} strong-coupling. These authors suggested that globally
imposed anisotropy might then increase the stability of the $A$-like phase. Aoyama and
Ikeda\cite{Aoy06} explored this idea theoretically finding that this might indeed be the case and
that one might expect a difference between samples that were axially compressed, as opposed to
radially compressed; the latter possibly stabilizing the polar state in preference to the axial
state.  The first experiments with globally induced axial anisotropy were reported by Davis {\it
et al.}\cite{Dav06} where radial compression of 10\% was induced by preferential shrinkage during
growth and drying of the aerogel. In this work a stable region of superfluid was indeed established,
although the nature of this phase has not been explored. In later work Davis {\it et al.}\cite{Dav08}
studied a 17\% axially compressed aerogel, also nominally 98\% in porosity, using transverse acoustic
impedance methods to detect the $A$-like to $B$-phase transition.  This sample was also characterized
by optical birefringence techniques.\cite{Pol08,Pol08LT25}  The results of the
supercooling experiments are compared in Fig.~20 with those for nominally unstrained samples used by
Gervais {\it et al.}\cite{Ger02} and Nazaretski {\it et al.},\cite{Naz04} both in zero applied
magnetic field. The supercooling region was approximately the same in all three cases at high
pressure. But at low pressure, $P \sim 10$ bar there was significantly more supercooling observed in
the axially compressed sample. Davis {\it et al.}\cite{Dav06} did not find evidence for a stable
region of $A$-like phase near $T_{ca}$ in their strained sample.

\begin{figure}[b]
%%%%%%%%%%%%%%%%%   F I G U R E  20   %%%%%%%%%%%%%%%%%%
\centerline{\includegraphics[width=2.8in]{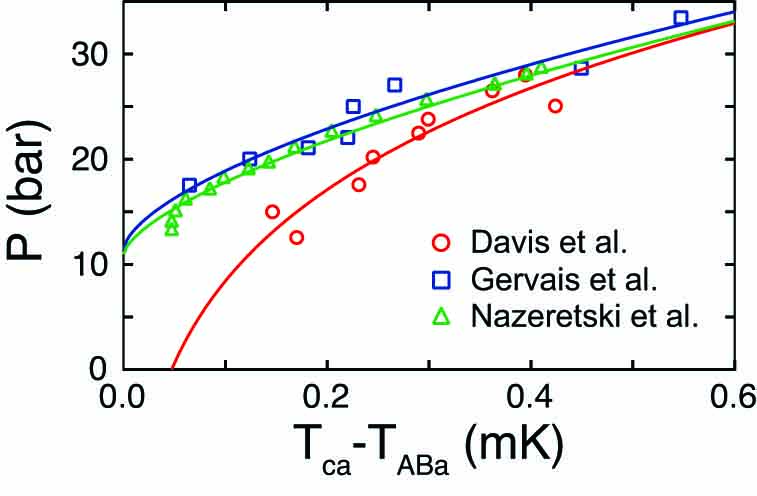}}
%%%%%%%%%%%%%%%%%%%%%%%%%%%%%%%%%%%%%%%%%%%%%
\begin{minipage}{0.93\hsize}
\caption {Supercooling of the metastable $A$-like phase as reported for a 17\% axially strained
sample of 98\% aerogel by Davis {\it et al.}\cite{Dav08} as compared with nominally unstrained
samples from Gervais{\it et al.}\cite{Ger02} and Nazaretski {\it et al.}\cite{Naz04} The extent of
supercooling is approximately the same.  Additionally, Davis {\it et al.}\cite{Dav08} found no
evidence for a stable region of $A$-like phase near $T_{ca}$.}
\end{minipage}
\end{figure}
\noindent

Most NMR experiments report that the spectrum of $^3$He in aerogel becomes strongly inhomogeneously
broadened in the superfluid state.  This phenomenon, in contrast to what is observed in
similar size samples of pure $^3$He, has its origin in a distribution of orientations of the order
parameter, or textures.\cite{Vol90}  It is likely that the texture distributions can be attributed to
ansiotropy in the aerogel.  In a remarkable experiment, Kunimatsu {\it et al.}\cite{Kun07} showed
that the NMR spectrum in one sample was maximally shifted to negative values and they
interpreted this phenomenon to alignment of the angular momentum
vector, $\bf \ell$, with the anisotropy axis and parallel to the magnetic field.  The existence of
significant axial anisotropy in that sample was later confirmed by optical birefringence
measurments performed at Northwestern. In two other NMR experiments, Dmitriev {\it et
al.}\cite{Dmi08} and Elbs {\it et al.}\cite{Elb08} have shown unusual order parameter textures
which they identified with  anisotropy within the aerogel. A common feature of the two
experiments is the coexistence of both
$B$ and
$A$-like phases for a range of temperature and clear evidence for the orientation of
$\bf \ell$ by the aerogel structure. It is interesting that the NMR spectra from the radially
compressed aerogel of Elbs {\it et al.}\cite{Elb08} are similar to the spectra obtained for
unstrained samples reported by Nakagawa {\it et al.}\cite{Nak05}

\section{Conclusions}

In the past 12 years, considerable experimental and theoretical work  has been reported  on the
effects of highly porous silica aerogel on the superfluid states of $^3$He.  The aerogel structure
provides a distribution of point-like scattering centers that can be interpreted as impurities.  In
this review we have concentrated on experiments and theoretical predictions of the scattering
models and we find that there is very good agreement between homogeneous scattering models and
experiments except for the transition temperature. In order to account for both the transition
temperature suppression and the other properties of superfluid $^3$He some form of inhomogeneous
scattering must be present.  The phenomenological inhomogeneous scattering model of Sauls and
Sharma meets this requirement with two parameters, the transport mean
free path and the aerogel particle-particle correlation length.  As a result, a consistent
description in terms of these two parameters can be obtained for the phase diagram, specific
heat, thermal conductivity, magnetic susceptibility, superfluid density, NMR longitudinal resonance
frequency, and the high field $A_{1}$ - $A_{2}$ splitting of the normal to superfluid transition. 

Exciting work currently in progress is exploring the possibility of manipulating the superfluid
state using specially prepared aerogel samples that exhibit global anisotropy with the broad view
that such anisotropic materials may stabilize anisotropic superfluid states, possibly states that
otherwise would not exist.  It is already clear from NMR experiments that the orientation of the
order parameter in the metastable $A$-like phase can be affected by global anisotropy.

There remains an unresolved problem of the nature of the metastable $A$-like phase and the related
question of superfluid-phase nucleation, particularly within a very narrow region close to the
transition temperature from normal to superfluid.

	We acknowledge support from the National Science Foundation,  DMR-0703656 and thank W.J. Gannon
for useful discussions.

\end{document}